# Role of Cybersecurity and Blockchain in Battlefield of Things

Gaurav Sharma[a], Deepak Kumar Sharma[b], Adarsh Kumar[c,*]

[a,b,c] *School of Computer Science, University of Petroleum and Energy Studies, Dehradun, India*
[a]*gauravsharmaiaf@gmail.com,* [b]*dksharma@ddn.upes.ac.in,* [c]*adarsh.kumar@ddn.upes.ac.in*

**Abstract**

The Internet of Things is an essential component in the growth of an ecosystem that enables quick and precise judgments to be made for communication on the battleground. The usage of the battlefield of things (BoT) is, however, subject to several restrictions for a variety of reasons. There is a potential for instances of replay, data manipulation, breaches of privacy, and other similar occurrences. As a direct result of this, the implementation of a security mechanism to protect the communication that occurs within BoT has turned into an absolute requirement. To this aim, we propose a blockchain-based solution that is both safe and private for use in communications inside the BoT ecosystem. In addition, research is conducted on the benefits of integrating blockchain technology and cybersecurity into BoT application implementations. This work elaborates on the importance of integrating cybersecurity and blockchain-based tools, techniques and methodologies for BoT.

## 1. Introduction

The battlefield in the present generation and in the times to come will be very complex as it will be an integration of networks wherein different sensors will be pumping information across different bandwidths. The person who utilizes the electromagnetic spectrum most judicially will emerge as the winner. As the world is moving towards the Internet of Things, the battlefield is also progressing toward the concept of the Internet of Battlefield of Things (IBoT). The focus of Battlefield of Things (BoT) is to provide situational awareness of the battlefield utilizing a network of interconnected sensors, actuators, and analytical devices [1]. Sensors will be able to detect the movement of the adversary and the relay mechanism will help relay the real-time location/ position of the enemy. The tactical decision of positioning the resources, the areas that need to be avoided, and the decision of where, when and which weapon platform to use will be governed by this tactical update. Cyber Security for IoT environments refers to mechanisms used to safeguard objects from physical damage, unauthorized access, information theft, data loss, and information about the thing whenever needed. The most important factor to be considered while designing a BoT is the security and confidentiality of the data that is being transmitted for military communication. Hence the importance of cybersecurity while designing a BoT scenario becomes vital. With the advancements in blockchain technology, the security and availability of information for a BoT environment can be best utilized for BoT environments [2-3]. Various applications and technologies of BoT are briefly explained as follows.

- **New dimension to Network-centric warfare (NCW):** The NCW has considerably matured over time in all the defence forces. However, the biggest drawback of NCW is its dependency on vintage radars and systems with considerable latencies. These are a drawback to quick decision making. Integrating the sensors with minimal latencies will add a new dimension to the NCW. The cloud-based architecture of the BoT will aid in larger databases, in-depth analysis, and quick decision making.
- **Increased situational awareness**: The IoBT will integrate multiple sensors and thus there will be a significant increase in the number of input sources. Also, since the technology is aided by cloud servers, there will be more data for analytics which leads to better situational awareness.
- **Quick decision making**. The loop for decision making is quite elaborate and foolproof. However, the decision making relies upon the input from the sensors. The limitations of the technology of these sensors lead to dead time which varies from sensor to sensor. With IoBT in place, the decision could be taken in real-time and thus would enhance the warfighting capability to a great extent.
- **Data Analytics**: Analytical methods and tools need to be incorporated to extract meaningful information from the data that has been collected.
- **Scalability**: In a battlefield aided by IoBT technology, there will be a considerable increase in the number of sensors that can be integrated. Thus the scalability factor is going to increase manifold.
- **Fault Tolerance**: In the present scenario, the decision makers have input from different sensors based on which the decision is made. However, if a particular sensor goes down, the decision has to be made with the remaining sensors until the sensor comes up again. With IoBT the algorithms will be able to develop a fault tolerant system, where even if a sensor goes down the data can either be calculated based upon data analytics or another pre designated entity to replace it.

- **Fleet Monitoring**: Fleet monitoring in the military refers to maintaining and managing aircraft/ ships/ equipment of a similar category or nature. Fleet monitoring ensures that the equipment is kept warfighting worthy at all times and is ready to undertake the mission as desired. It monitors the health status, serviceability, downtime measurement, and predictability for availability for a specific time duration.
- **Sensor utilization**: The BoT will redefine the process of sensor utilization in the military environment. Presently a single sensor gives feed/ input to the soldier on the battlefield in his area of operation. With BoT, the sensor input would be received at a central location like a private cloud and continuous monitoring up to the last sensor on a real time basis could be ensured.
- **C4ISR (Command Control Computer Communication Surveillance and Reconnaissance)**: The C4ISR technologies give command and control abilities, and electronic surveillance capabilities and help control and guide the fighter aircraft in a combat situation. With BoT, the C4ISR technologies will transform modern-day warfare. The fighter airplanes will be able to receive input from multiple sensors and will be monitored from a central location by IoT sensors integrated with the C4ISR aircraft.
- **UAV**: The most important war-fighting weapons of the future would be drones. The Drones have the advantage of being unmanned and can be controlled from a remote location. With the networking of drones, the drones can communicate to very large ranges. The drones come in different categories based on their roles like surveillance, monitoring and weapon delivery [4-5]. With BoT the drone technology will be able to perform far beyond its present capabilities. The swarm of drones concept can be best achieved in a BoT scenario [5].
- **Medicine:** The military has a large medical corps that caters to both peace time and war time requirements of the military personnel and their families. With IoT technologies, the medical corps of the military can benefit from the feasibility of accessing the medical records of its personnel through electronic means. This would become very useful in an emergency scenario on a battlefield where generally the medical records of the patient are not available.

## 1.1. Work Contributions

The major contributions of this work are briefly explained as follows.
- To identify the BoT, its infrastructure requirements and security aspects.
- To associate and elaborate on the importance of cybersecurity for BoT from futuristic perspectives.
- To identify the security aspects that can be achievable through blockchain and its associated terminologies for BoT cyberspace.

## 1.2. Organization of Work

This work is organized as follows. Section 1 introduces the importance of BoT. Section 2 explains the role and importance of cybersecurity in BoT. Further, this section explains the important tools and techniques for BoT. Section 3 shows the importance of blockchain in BoT. Further, a use case presenting blockchain for BoT is discussed. Finally conclusion is drawn in section 4.

## 2. Cybersecurity for BoT

In the present era, every computational device is interconnected through the Internet, or networks and communicates from one device to another device exchanging information digitally from source to destination. The BoT does every task or operation smoothly and properly under the timeline. Also, the BoT devices connect with WiFi, sensors, and Bluetooth network and send information command centre to the base station related mission "information" it's confidential or secret, so cyber security is a most important role in the BoT [6][7][8][9]. A detail of how the secure communication system works in the BoT network is shown in fig 1.

*Cyber security for BoT:* Cyber security refers to employing various technologies, procedures, and controls to protect BoT systems, networks, applications, devices, and data from being compromised by malicious cyber attackers. Its goal is to reduce the likelihood of cyberattacks and to prevent unwanted use of computer systems, networks, and other technological resources. It is divided into three general categories: information security, application security, and network security [10].

- **Information Security:** Information security prevents unauthorized access to, modification of, recording of, or interruption of any sensitive information. The ultimate objective is to protect sensitive data, such as mission information, artillery and missile, the number of fighter planes, and other battlefield network details. Theft of personal information, tampering with data, and data deletion are possible outcomes of security breaches. Attacks can cause disruption and damage to army defence operations; the three foundations of information security are confidentiality, integrity, and availability [11][12].
- **Application Security:** The objective of application security is to secure software application code and data from cyber threats. Application security may and should be applied across all phases of development, including design, development, and deployment. Include vulnerability assessment throughout the early phases of development and conduct continuous security testing; Implement strong authentication for mission-critical or sensitive data-containing applications [13].

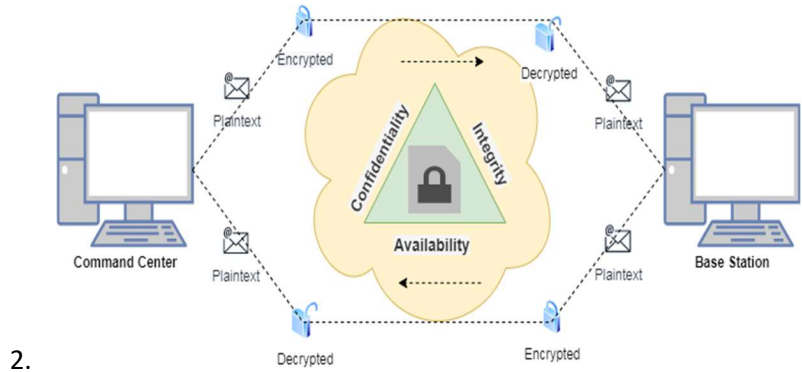

Fig: 1. Integrated Secure BoT networks

• **Network Security**: Network security is the technique of securing a BoT device network against intruders, whether targeted attackers or opportunistic malware, and refers to the practice of protecting a network from unauthorized access [13].

*Cyber Security Tools and Techniques for BoT*: Detailed tools and techniques used to secure the IBoT network are shown in table 1. Details include how to protect the BoT network's information from cyber-attacks and how can we prepare to be safe from future cyber attacks.

Table: 1. Cyber Security Tools and Techniques for BoT.

| Name of Tools | Major Techniques | Description |
|---|---|---|
| Mirai Botnet [9] | Distributed Denial-of-Service Attacks | During the investigation, a botnet comprising infected printers, IP cameras, residential gateways, baby monitors, and other internet-connected devices was discovered. |
| Muhstik Botnet [10] | Web application exploits | The Muhstik botnet frequently targets online application vulnerabilities to compromise IoT devices, such as cloud security enterprise Lacework for Drupal and Oracle WebLogic. |
| Xbash Botnet [11] | Ransomware | Xbash Bonet is data-destructive and hinders Linux databases. There is also no way to restore the Xbash after paying the ransom. |
| Dark Nexus [12] | Distributed Denial of Service attacks | Dark Nexus may assist in spam and phishing attacks but is typically devoted to destructive DDoS or Distributed Denial of Service attacks on the Internet. |
| Mozi [13] | Distributed denial-of-service (DDoS) attacks | Mozi is well-known for impacting maliciously IoT devices worldwide. The Mozi malware contains source codes from Mirai, IoT Reaper, and Gafgyt, and it joins a peer-to-peer (P2P) botnet and communicates with other infected host devices. |

## 3. Blockchain for BoT

The BoT (BoT) is becoming an essential part of military applications. These applications connect tanks, battlefield vehicles, warfighters, and drones so that they can work together to complete the objective. For tactical command and control, the objective of the BoT is a dynamic, flexible, and goal-driven networked battlefield environment [19-20]. In the future, smart things and machine intelligence will replace human authorities and war combatants in military operations. The BoT will connect warfighters with advanced technology to improve armor, weapons, and other things to provide warfighters with enhanced sensory awareness, situational knowledge, prediction power in critical situations, and risk assessment. The opponents, on the other hand, may be able to subvert the operations and directives that come from higher authorities or peers, which may cause the entire mission to become a failure. Therefore, having communications that are reliable and cannot be altered is necessary for BoT [21]. Blockchain technology should be used to overcome these difficulties, and improve IoBT device information trustworthiness. Blockchain creates a peer-to-peer network of immutable distributed ledgers for IoBT devices to connect. Blockchain is excellent for military missions where multiple devices are connected and communicate in a network. Blockchain is transparent and cost-effective making it appealing for IoBT devices. A decentralised distributed digital ledger, blockchain, is a consensus system that is considered to bring a new era of applications. Trusted third parties are not required to implement this system. According to its description, a blockchain is a distributed ledger that has the capability of recording transactions between two parties in a form that is both verifiable and permanent. A blockchain is a peer-to-peer network that connects all of its nodes and verifies and adds new blocks to the network. BoT is the result of applying blockchain technology to military operations, and it enables trusted, informed decisions to be made by the military[20-21].

***Advantages and challenges of blockchain-based IoBT :*** The whole concept of BoT is based on the fact that confidentiality, integrity and availability of data will not be compromised come what may. Thus, having a strong secure and robust mechanism to safeguard the military data that is either stored or communicated over the network becomes paramount. Here Blockchain technology comes to the rescue. There are many advantages of using Block Chain in BoT. Some of the advantages of using blockchain in the IoBT communication environment are [3,18]: Immutability, availability, no single point of failure, data transparency and traceability, safe data transfer, and secure data storage. There is a significant number of challenges in blockchain-based IoBT. Some of these challenges are [3,20]: Operational Cost, Time complexity of adding data, Scalability and Data Privacy.

***Blockchain-based model for BoT:*** Blockchain-based IoBT generic model is shown in figure 2 which contains three important layers: the internet of battlefield layer, network layer, and consensus layer [2,20].

- **Internet of Battlefield Layer:** The internet of battlefield layer consists of all military objects such as tanks, helicopters, drones, troops etc. that sense and track adversary activity in physical conflict space. Mission success depends on environmental sensing, troop progress monitoring, inventory management, and soldier health checks. With this information, timely support can be arranged, reducing the risk of mission failure [22].
- **Network Layer:** From the IoBT, nodes capture valid transactions and send them to the closest Blockchain nodes which are responsible for consensus. Usually, some of the nodes are full nodes and other nodes may act as endorsers. The endorser node is responsible for verifying the transaction's validity and authenticity and further sending the status to full nodes. Full nodes take the help of endorser nodes for verifying transactions [23-24].
- **Consensus Layer:** An important role of this consensus layer is to use standard agreement processes to accept transactions from BoT . The nodes in the network layers receive transactions that check the validity. A block is created for each legitimate transaction, which is then updated into the distributed ledger after the network reaches consensus. In the case of off-chain blockchain, data is stored and managed by either cloud storage or distributed file system. A suitable consensus mechanism is selected for cost-effective and reliable blockchain-based BoT. There are various lightweight consensus mechanisms for BoT[25], and it seems that the best-suited one can be resource-dependent (Proof-of-Work). However, the major blockchain networks are preferring proof-of-stake compared to proof-of-work. Lightweight consensus algorithms provide a reduced cost in terms of communication and computation and high security with lightweight cryptographic primitives and protocols. This adds reliability as well. Thus, lightweight data consensus algorithms would be an efficient approach to ensure low-cost and high security-based approach for BoT[26][27][28][29].
- **Smart Contracts within Blockchain**: Smart contracts are computer programs that are an automated mechanism of transfer of digital assets between various stakeholders. Smart contracts are enabled with blockchain to provide adequate security while dealing with cryptocurrency. Each stakeholder has a copy of the smart contract and it cannot be changed by any single stakeholder. The outcome of the smart contract will be the same irrespective of who executes the contract since it is digitally programmed. There is no third-party involvement in the case of a smart contract, in case a smart contract is executed between two parties, only they design it and have access to it. It is customizable, transparent and accurate.

***Use cases for Blockchain-based BoT:***

**Use of Blockchain in Missile Defence System**: Let us consider a use case for use of a blockchain based missile defence system. In able to understand the use of blockchain it will be relevant to understand the existing missile defence system of any typical military.

(i) **Present Typical Missile Defence system**. There are various sensors that will be gathering signal intelligence and monitoring friendly and enemy air space. If an enemy aircraft is detected by any one sensor, the information about the enemy aircraft is relayed to the command and control centre. The command and control centre then analyses this information by correlating it with the inputs received from other sensors as well. Once the information has been scrutinized and authenticated, it relays the information of one of its missile command posts to intercept the enemy aircraft by launch of a missile.

(ii) **Missile defence system based on Blockchain technology**. In the case of a blockchain based missile defense system, as soon as the sensor picks up a signal of an enemy aircraft intruding into the air space, the sensor communicates this using blockchain to the central command and control centre. This communication is encrypted and digitally signed. Thus there is no scope that this information can be either interpreted by the enemy or falsified information relayed to the command centre due to enemy hacking into the network. Also as approving the transaction requires validation (In a regular scenario, it is part of the consensus mechanism defined within the consensus layer. However, if the sequence of activities is to be decided by nodes then they may take an explicit call as well. ) from each and every node, the enemy will have to actually hack into each and every node which is practically impossible.

Each node in this scenario is protected by an encryption mechanism with strong cryptography algorithms. Thus practically impossible to be hacked. Further after ascertaining the threat, the command and control centre orders the missile post to launch a missile onto the enemy aircraft.

**Managing Military Logistics through Blockchain.** The benefits of blockchain technology in military logistics and supply chain and result in assured delivery of military goods, safety, traceability and reduced costs. Military fights its battles based upon efficient backing of its logistics support both for warfighting reserves and also the supply of human resources. All this supply chain can be automated by the use of blockchain technology. One of the greatest advantages of blockchain in military logistics will be that the supplies will be able to be tracked by each and every one. So not only will the soldier on the ground will be able to know when the logistics are reaching him/her, even the military commander will be able to track the consignment on a real-time basis. Also, all these transactions will be foolproof and secure as they will be encrypted and will use digital signatures so that the enemy cannot hack and mislead the troops about regular military supplies.

**The Russia-Ukraine War Lessons learnt from a cybersecurity perspective:** Figure 3 shows an example of hierarchical technological-integrated Battlefield of Things operational architecture. When Russia invaded Ukraine on 24 Feb 2022, large-scale multi-level attacks to bring down the cyber domain of Ukraine were reported wherein multi-state actors including hackers, state-sponsored cyber professionals, and malicious attackers targeted areas of varied interests. The war brought in the realization among leading nations about the Internet of Battlefield of Things. A major reason for this was the fact that the domain of war itself had changed when compared to other traditional wars of the past. The Ukrainians had cell phones and access to social media platforms which on one hand removed the fog of war but on other hand made the forces fighting the war on the ground vulnerable to being tracked and attacked by the Russian forces. The cyber-war which is the key factor in the Russia-Ukraine war covers the entire gambit of network-centric operations which includes the internet, intranet and the physical infrastructure that supports the architecture. One of the major players from the perspective of IoBT in Russia Ukraine War was the use of Drones by either side. In conclusion, the important technologies helpful in the modern battlefield include but are not limited to IoBT for secure and efficient connectivity for communication, blockchain for security and authentication of data (at all stages including storage, processing and transmission) and users, cloud services for large scale and distant connectivity.

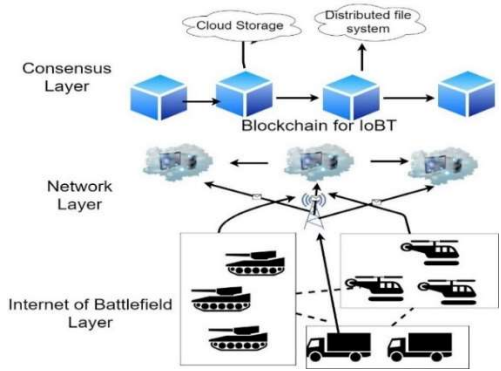

Fig: 2. Blockchain based IoBT model

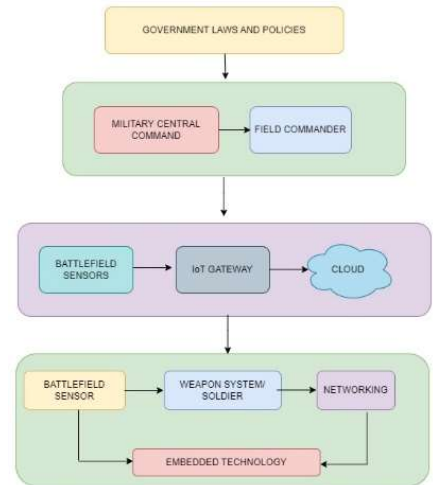

**Figure 3:** Hierarchical Technological-integrated Battlefield of Things Operational Architecture

## 4. Conclusion

The IoBT environment comprises of devices that are both smart and intelligent. The communication environment provided by IoBT is one that is excellently suited for the purpose of battlefield communication. However, an environment like this is susceptible to a variety of various kinds of attacks. In this article, we have discussed the role of cybersecurity along with cybersecurity tools and technologies in BoT. We also discussed the role of blockchain for BoT that ensures immutability, privacy, and a reliable environment. A generic model for the blockchain-based BoT has been proposed for efficient communication between various devices.

**Data Availability**
There is no associated data related to this article.
**Conflicts of Interest**
The authors declare that they have no potential conflict of interest.
**Funding Statement**
This research did not receive any specific grants from public, commercial, or non-profit funding bodies.

## References

[1] Varghese, Vincent, S. Sundeep Desai, and Manisha J. Nene. Decision making in the battlefield-of-things. *Wireless Personal Communications* 106, no. 2 (2019): 423-438. https://doi.org/10.1007/s11277-019-06170-y

[2] D. K. Tosh, S. Shetty, P. Foytik, L. Njilla and C. A. Kamhoua. Blockchain-Empowered Secure Internet -of- Battlefield Things (IoBT) Architecture. MILCOM 2018 - 2018 IEEE Military Communications Conference (MILCOM), 2018, pp. 593-598, doi: 10.1109/MILCOM.2018.8599758.

[3] T. Aste, P. Tasca and T. Di Matteo. Blockchain Technologies: The Foreseeable Impact on Society and Industry. *Computer*, vol. 50, no. 9, pp. 18-28, 2017. DOI: 10.1109/MC.2017.3571064

[4] Sharma, K., Singh, H., Sharma, D.K., Kumar, A., Nayyar, A., Krishnamurthi, R. (2021). Dynamic Models and Control Techniques for Drone Delivery of Medications and Other Healthcare Items in COVID-19 Hotspots. In: Al-Turjman, F., Devi, A., Nayyar, A. (eds) *Emerging Technologies for Battling Covid-19*. Studies in Systems, Decision and Control, vol 324. Springer, Cham. https://doi.org/10.1007/978-3-030-60039-6_1

[5] Kumar, A., Bhatia, S., Kaushik, K., Gandhi, S.M., Devi, S.G., Diego, A.D.J. and Mashat, A., 2021. *Survey of promising technologies for quantum drones and networks*. IEEE Access, 9, pp.125868-125911

[6]. Zhu, L., Majumdar, S., & Ekenna, C. (2021). An invisible warfare with the internet of battlefield things: a literature review. *Human behavior and emerging technologies*, *3*(2), 255-260.

[7]. Seo, S., Han, S., & Kim, D. (2022). D-CEWS: DEVS-Based Cyber-Electronic Warfare M&S Framework for Enhanced Communication Effectiveness Analysis in Battlefield. *Sensors*, *22*(9), 3147.

[8] Jain, S., Tomar, D. S., & Sahu, D. R. (2012). Detection of javascript vulnerability at Client Agen. *International Journal of Scientific & Technology Research*, 1(7), 36-41.

[9] Basit, Z., Tabassum, M., Sharma, T., Furqan, M., & Md, A. Q. (2022). Performance analysis of OSPF and EIGRP convergence through IPsec tunnel using Multi-homing BGP connection. *Materials Today: Proceedings*. https://doi.org/10.1016/j.matpr.2022.03.486

[10] Latif, S. A., Wen, F. B. X., Iwendi, C., Li-li, F. W., Mohsin, S. M., Han, Z., & Band, S. S. (2022). AI-empowered, blockchain and SDN integrated security architecture for IoT network of cyber physical systems. *Computer Communications*, 181, 274-283. https://doi.org/10.1016/j.comcom.2021.09.029

[11] Mohd, N., Singh, A., Bhadauria, H. S., & Wazid, M. (2022). An efficient node placement scheme to mitigate routing attacks in Internet of Battlefield Things. *Computers & Electrical Engineering*, 97, 107623. https://doi.org/10.1016/j.compeleceng.2021.09.029

[12] Usman, M., Amin, R., Aldabbas, H., & Alouffi, B. (2022). Lightweight challenge-response authentication in SDN-based UAVs using elliptic curve cryptography. *Electronics*, *11*(7), 1026. https://doi.org/10.3390/electronics11071026

[13] Zibetti, G. R., Wickboldt, J. A., & de Freitas, E. P. (2022). Context-aware environment monitoring to support LPWAN-based battlefield applications. *Computer Communications*, *189*, 18-27. https://doi.org/10.1016/j.comcom.2022.02.020

[14] Waqas, M., Kumar, K., Laghari, A. A., Saeed, U., Rind, M. M., Shaikh, A. A., ... & Qazi, A. Q. (2022). Botnet attack detection in Internet of Things devices over cloud environment via machine learning. *Concurrency and Computation: Practice and Experience*, *34*(4), e6662. https://doi.org/10.1002/cpe.6662.

[15] de Melo, P. H., Miani, R. S., & Rosa, P. F. (2022). FamilyGuard: A Security Architecture for Anomaly Detection in Home Networks. *Sensors*, *22*(8), 2895. https://doi.org/10.3390/s22082895.

[16] Gangwar, S., & Narang, V. (2022). A Survey on Emerging Cyber Crimes and Their Impact Worldwide. In *Research Anthology on Combating Cyber-Aggression and Online Negativity* (pp. 1583-1595). IGI Global. DOI: 10.4018/978-1-6684-5594-4.ch080.

[17] Jove, E., Aveleira-Mata, J., Alaiz-Moretón, H., Casteleiro-Roca, J. L., Marcos del Blanco, D. Y., Zayas-Gato, F., & Calvo-Rolle, J. L. (2022). Intelligent One-Class Classifiers for the Development of an Intrusion Detection System: The MQTT Case Study. *Electronics*, *11*(3), 422. https://doi.org/10.3390/electronics11030422.

[18] Tu, T. F., Qin, J. W., Zhang, H., Chen, M., Xu, T., & Huang, Y. A comprehensive study of Mozi botnet. *International Journal of Intelligent Systems*. 2022. https://doi.org/10.1002/int.22866

[19] Adebayo, Abdulhamid, Danda B. Rawat, Laurent Njilla, and Charles A. Kamhoua. Blockchain-enabled information sharing framework for cybersecurity. *Blockchain for Distributed Systems Security* (2019): 143-158.

[20] M. Wazid, A. K. Das, S. Shetty and J. J. P. C. Rodrigues. On the Design of Secure Communication Framework for Blockchain-Based Internet of Intelligent Battlefield Things Environment. IEEE INFOCOM 2020 - IEEE Conference on Computer Communications Workshops (INFOCOM WKSHPS), 2020, pp. 888-893, doi: 10.1109/INFOCOMWKSHPS50562.2020.9163066.

[21] U. Khakurel, D. Rawat and L. Njilla. FastChain: Lightweight Blockchain with Sharding for Internet of Battlefield-Things in NS-3. 2019 IEEE International Conference on Industrial Internet (ICII), 2019, pp. 241-247, doi: 10.1109/ICII.2019.00050.

[22] Sharma, D.K., Kumar, A. and Bathla, G., 2022. Heavy vehicle defense procurement use cases and system design using blockchain technology. In Autonomous and Connected Heavy Vehicle Technology (pp. 287-301). Academic Press. https://doi.org/10.1016/B978-0-323-90592-3.00017-3

[23] Kumar, A., Pacheco, D., Kaushik, K., Rodrigues, J. J. P. C.. Futuristic View of the Internet of Quantum Drones: Review, Challenges and Research Agenda. Vehicular Communications, Elsevier, 100487, 2022, ISSN 2214-2096, https://doi.org/10.1016/j.vehcom.2022.100487.

[24] G. Bathla, K. Bhadane, R.K.Singh, R. Kumar, R. Aluvalu, R. Krishnamurthi, A. Kumar, R.N. Thakur and S. Basheery. Autonomous Vehicles and Intelligent Automation: Applications, Challenges, and Opportunities. *Mobile Information* Systems. 2022. https://doi.org/10.1155/2022/7632892 .

[25] Kumar, A., Kumar Sharma, D., Nayyar, A., Singh, S. and Yoon, B., 2020. Lightweight proof of game (LPoG): A proof of work (PoW)'s extended lightweight consensus algorithm for wearable kidneys. Sensors, 20(10), p.2868.

[26] Zhang, W., Wu, Z., Han, G., Feng, Y. and Shu, L., 2020. LDC: A lightweight dada consensus algorithm based on the blockchain for the industrial Internet of Things for smart city applications. Future Generation Computer Systems, 108, pp.574-582.

[27] Kumar, A. and Hedabou, M., 2022. An Uncertainty Trust Assessment Scheme for Trustworthy Partner Selection in Online Games. IEEE Access.

[28] Bentajer, A., Said, Y., Igarramen, Z. and Hedabou, M., 2021, November. Robust Secret Share to Reinforce the Security of IBE's Master Key. In The International Conference on Information, Communication & Cybersecurity (pp. 526-534). Springer, Cham.

[29] Kumar, A. and Sharma, D.K., 2021. An optimized multilayer outlier detection for internet of things (IoT) network as industry 4.0 automation and data exchange. In International Conference on Innovative Computing and Communications (pp. 571-584). Springer, Singapore.